\documentclass[twocolumn]{aa}  

\usepackage{xcolor}
\usepackage{bm}
\usepackage{amsmath}
\usepackage{graphicx}
\usepackage{txfonts}
\usepackage{dsfont}
\defcitealias{jaupart2021a}{Paper I}
\newcommand{\esp}[1]{\mathbb{E}\left( {#1} \right)}
\newcommand{\var}[1]{\mathrm{Var}\left( {#1} \right)}

\begin{document}

   \title{Statistical properties and correlation length in star-forming molecular clouds: II. Gravitational potential and virial parameter }
 \titlerunning{Statistical properties and correlation length: II.  Gravitational potential and virial}

   \author{E. Jaupart \inst{1} \and G. Chabrier\inst{1}$^,$ \inst{2} }
   \authorrunning{Jaupart, Chabrier}

   \institute{Ecole normale sup\'erieure de Lyon, CRAL, Universit\'e de Lyon, UMR CNRS 5574, F-69364 Lyon Cedex 07, France\\
              \email{etienne.jaupart@ens-lyon.fr}
         \and
             School of Physics, University of Exeter, Exeter, EX4 4QL, UK\\
             \email{chabrier@ens-lyon.fr} }

   \date{Received 14/04/2021; accepted 16/04/2022}

 
  \abstract{
  In the first article of this series, we have used the ergodic theory to  assess the validity of a statistical approach to characterize various
properties of star-forming molecular clouds (MCs) from a limited number of observations or simulations. This allows the proper determination of  confidence intervals for various volumetric averages of statistical quantities obtained
form observations or numerical simulations. We have shown that these confidence intervals, centered on the statistical average of the given quantity, decrease as the ratio of the correlation length to the  size of the sample gets smaller.

In this joint paper, we apply the same formalism to a different kind of (observational or numerical) study of MCs. Indeed, as observations cannot fully unravel the complexity of the inner density structure of star forming clouds, it is important to know whether global observable estimates, such as the total mass and size of the cloud, can give an accurate estimation of various key physical quantities that characterize the dynamics of the cloud. Of prime importance is the correct determination of the total gravitational (binding) energy and virial parameter of a cloud. We show that, whereas for clouds  that are not in a too advanced stage of star formation, such as Polaris or Orion B, the knowledge of only their mass and size  is sufficient to yield an accurate determination of the aforementioned quantities from observations (i.e. in  real space). In contrast, we show that this is no longer true for numerical simulations in a periodic box. We derive a relationship for the ratio of the virial parameter in these two respective cases.
}

 \keywords{methods: analytical, methods: statistical, ISM: clouds,  ISM: structures, ISM kinematics and dynamics}

   \maketitle
%
\section{Introduction}

In a previous article (\citealt{jaupart2021a}, hereafter Paper I) we  derived a general framework aimed at assessing the relevance and validity of a statistical approach to characterize various
properties of star-forming molecular clouds (MCs) from a limited number of observations or simulations. We calculated the auto-covariance function (ACF) and correlation length of density fields in MCs and provided a way to determine the correct statistical error bars on observed probability density functions (PDFs). Applying these results to two typical star-forming clouds, Polaris and Orion B, which display two different types of PDFs, we have shown that the ratio of the correlation length of density fluctuations over the size of the cloud is typically $l_c/L\lesssim 0.1$. This justifies the relevance of an approach that uses the  hypothesis of statistical homogeneity when exploring star-forming MC properties, notably the evolution of the PDF, as carried out in \citet{jaupart2020}.

In this joint article, we apply the same formalism to a different kind of (observational or numerical) study of MCs. Indeed, as observations cannot fully unravel the entire complexity of the inner structure of star-forming clouds, it is important to know whether global observable estimates, such as the total mass and size of the cloud, can be used to give an accurate estimation of various key global physical quantities that are supposed to characterize the dynamics of the cloud. Of prime importance, for instance, is the correct determination of the total gravitational (binding) energy of a cloud, and its virial parameter, from estimations of its total mass and size only. 

\section{Correlation length and gravitational binding energy} \label{sec:gravpotanygeom}

In  \citetalias{jaupart2021a}, we  relied on the correlation length to determine confidence intervals for the measured statistical quantities. 
Here, we follow similar lines of argument to assess the accuracy of estimates of cloud characteristics deduced from  global properties without any knowledge of the cloud internal structure. 
We focus on the cloud potential energy, noted as $|e_{G}|$, and virial parameter, noted  as $\alpha_{\rm vir}$, which can be deduced from the total mass, $M$, and size, $L$. 
This key issue was raised in particular by \citet{federrath2012,federrath2013}. In their numerical simulations of turbulent star-forming MCs, 
these authors found a large discrepancy between values of the virial parameter deduced from the cloud's global (average) characteristics and those measured directly from the numerical results. 
These authors concluded that  ‘‘this shows that comparing simple theoretical estimates of the virial parameter, solely based on the total mass, as a measure for $|e_{G}|$ [...], 
should be considered with great caution because such an estimate ignores the internal structure of the clouds.'' We show below that  the dependence of the total potential energy and virial parameter 
on the ‘‘internal structure'' can be assessed from the knowledge of the correlation length of density fluctuations within the cloud, $l_c(\rho)$. We further show that the problem
raised by \citet{federrath2012} is due primarily to artificial effects stemming from the resolution of the Poisson equation in a periodic box.

We first started by deriving the total potential energy, noted  as $e_{G}$\footnote{The usual definition of the binding energy involves a 1/2 multiplying factor in order to account for summations on interacting pairs, which we omit here for the sake of clarity.},  
of a statistically homogeneous cloud in a domain, $\Omega$, in any geometry, and isolate the contribution of the internal structure from the rest. 
For the sake of simplicity, we  assumed that the cloud possesses a center of symmetry, which we take as the origin, such that $\forall \bm{y} \in \Omega$, $-\bm{y} \in \Omega$:
\begin{eqnarray}
    \left< e_{G} \right> &=& \left< e_{G} (\rho) \right> = \frac{1}{|\Omega|} \int_\Omega \rho(\bm{x}) \, \Phi_{G}(\bm{x}) \mathrm{d} \bm{x}  \\
    &=&\frac{1}{|\Omega|} \int_\Omega \rho(\bm{x}) \int_\Omega \rho(\bm{x'}) \, \Phi_{\rm Green}(\bm{x}-\bm{x'}) \, \mathrm{d} \bm{x'} \, \mathrm{d} \bm{x}  \\
     &=& \left< e_{G} (\left<\rho\right>) \right>  +  \int_{\Omega^2} \frac{\delta \rho(\bm{x}) \delta \rho(\bm{x'})}{|\Omega|} \Phi_{\rm Green}(\bm{x}-\bm{x'}) \, \mathrm{d} \bm{x'} \, \mathrm{d} \bm{x} \nonumber\\
     && + 2 \left<\rho\right> \int_{\Omega^2} \frac{\delta \rho(\bm{x})}{|\Omega|} \Phi_{\rm Green}(\bm{x}-\bm{x'}) \, \mathrm{d} \bm{x'} \, \mathrm{d} \bm{x} \\
     &=& \left< e_{G} (\left<\rho\right>) \right> + I_C(\rho) + 2 \left<\rho\right> I_\delta(\rho),
\end{eqnarray}
where $|\Omega|=L^3$ is the volume of the $\Omega$ domain, $\Phi_{\rm Green}$ is Green's function of the gravitation potential,  $ \Phi_{G}$, which is parity invariant, $\left<\rho\right>$ is the (volumetric) average density of the cloud,
\begin{equation}
\left<\rho\right> = \frac{1}{|\Omega|}\int_\Omega \rho(\bm{x}) \mathrm{d} \bm{x} = M/|\Omega| ,
\end{equation}
 $\delta \rho = \rho - \left<\rho\right>$, and $ \left< e_{G} (\left<\rho\right>) \right>  $ is the potential energy of the cloud if it were strictly homogeneous:
\begin{equation}
\left< e_{G} (\left<\rho\right>) \right> = \frac{1}{|\Omega|} \int_\Omega \, \left<\rho\right> \Phi_{G}(\bm{x}) \mathrm{d} \bm{x}.
\end{equation}

Using the change of fields $(\bm{u},\bm{v}) = \varphi(\bm{x},\bm{x'}) =(\bm{x} -\bm{x'}, \bm{x} + \bm{x'})$ and using $\hat{C}_{\rho,L}$ to denote the biased ergodic estimator of the $\rho$ ACF (see \citetalias{jaupart2021a}), we  obtained
\begin{eqnarray}
   I_C(\rho)  \! \!  &=&   \! \! \int_{\varphi_1(\Omega)}  \! \!   \mathrm{d} \bm{u} \Phi_{\rm Green}(\bm{u}) \int_{\varphi_2^{\bm{u}}(\Omega)}  \frac{ \mathrm{d} \bm{v}}{8|\Omega|}  \delta\rho(\frac{\bm{u} + \bm{v}}{2}) \delta\rho(\frac{\bm{u} - \bm{v}}{2}) \nonumber\\
    &=& \int_{\varphi_1(\Omega)}  \mathrm{d} \bm{u} \Phi_{\rm Green}(\bm{u}) \,  \hat{C}_{\rho,L}(\bm{u}), 
 \end{eqnarray}
 \begin{eqnarray}
  I_\delta(\rho)  \! \! &=& \! \! \int_{\varphi_1(\Omega)}  \! \!  \mathrm{d} \bm{u} \Phi_{\rm Green}(\bm{u}) \int_{\varphi_2^{\bm{u}}(\Omega)}  \frac{ \mathrm{d} \bm{v}}{8|\Omega|}  \delta\rho(\frac{\bm{u} + \bm{v}}{2}) \nonumber \\
  &\ll& \frac{\left< e_{G} (\left<\rho\right>) \right>}{2 \left< \rho \right>}, \label{eq:inequalitynegvirial}
\end{eqnarray}
where $(\varphi_1(\Omega),\varphi_2^{\bm{u}}(\Omega))$ is a parameterisation of $\varphi(\Omega^2)$  (see Appendix~\ref{app:ergodiccalculationgeneralvolume}). 
For example, 
\begin{eqnarray}
\mathrm{if} \, \,  \Omega &=& \left[-\frac{L}{2},\frac{L}{2}\right]^3, \\
\varphi_1(\Omega) &=&\left[-L,L\right]^3 \\
\mathrm{and} \, \,  \varphi_2^{\bm{u}}(\Omega)) &=& \left[-L+ |u_i|, L-|u_i|] \right]. 
\end{eqnarray}
We then write $ \hat{C}_{\rho} = \mathrm{Var}(\rho) \times \Tilde{C}_{\rho}$, such that $\Tilde{C}_{\rho}(0) = 1$ to finally obtain:
\begin{equation}
\left< e_{G} \right> \simeq \left< e_{G} (\left<\rho\right>) \right> + \var{\rho} \int_{\varphi_1(\Omega)}  \mathrm{d} \bm{u} \Phi_{\rm Green}(\bm{u}) \,  \Tilde{C}_{\rho,L}(\bm{u}). 
\end{equation}
In this expression, the contributions from the global (average) observables and the internal structure are separated explicitly. 
We now compare the case of two geometrical configurations, the real space $\mathbb{R}^3$,  relevant to observations, and the periodic simulation box $\mathbb{T}^3$.

\subsection{Isolated cloud} \label{sec:Isolated}

In $\mathbb{R}^3$ we have $\Phi_{\rm Green}(\bm{x})= - G/|\bm{x}|$; hence,
\begin{eqnarray}
    \left< e_{G} \right>_{\mathbb{R}^3} &=& \left< e_{G} (\left<\rho\right>) \right>_{\mathbb{R}^3} - G \var{\rho} \int_{\varphi_1(\Omega)}  \mathrm{d} \bm{u} 
    \,  \frac{\Tilde{C}_{\rho,L}(\bm{u})}{|\bm{u}|},
\end{eqnarray}
with 
\begin{eqnarray}
\left< e_{G} (\left<\rho\right>) \right>_{\mathbb{R}^3} = - 2 \, G \,  M c_g \frac{\left<\rho\right>}{L}, \label{eq:eprhomean}
\end{eqnarray}
where $c_g$ is a geometric factor of order unity if the cloud is roughly of the same dimension $L$ in the three directions. For example if $\Omega=B(R)$ is a ball\footnote{We recall that we omitted the usual factor 1/2 in the definition of the binding energy.}  of radius $R=L/2$,  $c_g=1.2$ while if $\Omega=\left[-\frac{L}{2},\frac{L}{2}\right]^3$ is a cuboid of size $L$, $c_g \simeq 1.9/2 \simeq 0.95$. Then,
\begin{eqnarray}
   \int_{\varphi_1(\Omega)}  \mathrm{d} \bm{u} 
    \,  \frac{\Tilde{C}_{\rho,L}(\bm{u})}{|\bm{u}|} = 8 \, \Tilde{c_g} \, l_c(\rho)^2 f\left(R/l_c(\rho)\right),
\end{eqnarray}
where $\Tilde{c_g}$ is a geometric factor of order unity and $f$ some function that converges rapidly towards $1$. For an exponential ACF  (see \citetalias{jaupart2021a}),  $\Tilde{c_g} = \pi^{1/3}/ 2 \simeq 0.73$ and $f(x) \simeq 1-(1+x) \mathrm{e}^{-x} $. As we have $l_{\rm c}/R \lesssim 0.1$ (see \citetalias{jaupart2021a}), we can write $f\left(R/l_c(\rho) \right)\simeq 1$ and
\begin{eqnarray}
    \left< e_{G} \right>_{\mathbb{R}^3} &\simeq& - 2 \, G \,  M c_g \frac{\left<\rho\right>}{L} \left(1 + \frac{4 \var{\rho} L}{  \left<\rho\right> M} \frac{\Tilde{c_g}}{c_g} l_c(\rho)^2 \right) \\
    &\simeq& - 2 \, G \,  M c_g \frac{\left<\rho\right>}{L} \left(1 + 2 \frac{\Tilde{c_g}}{c_g \, \xi_g} \var{\frac{\rho}{\left< \rho\right>}}  \left(\frac{l_c(\rho)}{R}\right)^2 \right), \label{eq:internalstructureVariance} 
\end{eqnarray}
where, $\xi_g$ is of order unity (for a ball $\xi_g= \pi/3$ and for a cube $\xi_g= 2$). In \citetalias{jaupart2021a} we derived a useful relation between the variance of the density field, $\rho$, the column density, $\Sigma$, and the correlation length, $l_c(\rho)$,  namely:
\begin{eqnarray}
\var{\frac{\Sigma}{\left< \Sigma\right>}} \simeq  \var{\frac{\rho}{\left< \rho\right>}} \frac{l_c(\rho)}{R},
\end{eqnarray}
providing that $l_c/R \ll 1$ (see Eq.~(44) in \citetalias{jaupart2021a}). Thus, we have
\begin{eqnarray}
\left< e_{G} \right>_{\mathbb{R}^3} &\simeq&  - 2 \, G \,  M c_g \frac{\left<\rho\right>}{L} \left( 1 + 2 \frac{\Tilde{c_g}}{c_g \, \xi_g}  \var{\frac{\Sigma}{\left< \Sigma\right>}}   \left(\frac{l_c(\rho)}{R}\right)  \right).\label{eq:internalstructureVariancObserv}
\end{eqnarray}
Eq.~(\ref{eq:internalstructureVariancObserv}) enables us to infer  the influence of the internal structure of the cloud from observations of column densities. We thus see that, if the product $\var{\rho/\left<\rho\right>}\times (l_c(\rho)/R)^2 \ll 1 $ for volume densities, or if the product $\var{\Sigma/\left<\Sigma\right>} \times (l_c(\rho)/R) \ll 1 $ for column densities, the correction of the cloud's internal structure contribution to the average gravitational energy is negligible. 

\subsubsection{Isothermal compressible turbulent conditions.}
For isothermal turbulent conditions, argued to be representative of initial conditions in star-forming clouds (see, e.g., \citealt{mckee2007} and reference therein), $\mathrm{Var}(\rho) \simeq (b \mathcal{M}) ^2 \esp{ \rho }^2 \simeq (b \mathcal{M}) ^2 \left< \rho\right>^2$ and we get 
\begin{eqnarray}
\var{\frac{\rho}{\left< \rho \right>}} \left(\frac{l_c(\rho)}{R}\right)^2 \simeq \left(\frac{(b \mathcal{M})l_c(\rho)}{R}\right)^2,
\end{eqnarray}
 where $b$ is the coefficient reflecting the driving mode contributions of the turbulence (\citealt{federrath2008}). For typical MC conditions in the Milky Way, $(b \mathcal{M}) \lesssim 5$ ($\mathcal{M} \sim 10$ and $b\simeq 0.5$). 
 
Moreover, in \citet{jaupart2021generalized}, we show that for compressible turbulence without gravity, within a factor of order unity,  $l_c(\rho) \sim \lambda_s \simeq L/\mathcal{M}^2$, where $\lambda_s$ is the sonic scale that is found to be close to the average width of filamentary structures in isothermal turbulence \citep{federrath2016}. However, as mentioned in the above article, while in case of pure gravitation-less turbulence the correlation length should be about the sonic length, this is not necessarily the case if gravity initially plays a non-negligible role.

In any case, since  as shown in \citetalias{jaupart2021a}, $l_{\rm c}(\rho)/R \lesssim 0.1$ (and more likely $l_{\rm c}(\rho)/R \sim 10^{-2}$)  with $\var{\Sigma/\left< \Sigma\right>} \lesssim 1$  in these clouds,  we conclude that 
 \begin{eqnarray}
    \left< e_{G} \right>_{\mathbb{R}^3} &\simeq&  - 2 \, G \,  M c_g \frac{\left<\rho\right>}{L} \left(1 + \Tilde{\xi_g} \right),
\end{eqnarray}
where $\Tilde{\xi_g}\equiv\Tilde{\xi_g}\left(b\mathcal{M}, l_c(\rho)/R\right)$  is again  at most of order unity for large values of $b{\cal M}\sim 10$ but more generally is on the order of a few percent. 

\subsubsection{Late-stage evolution of star-forming clouds}

However, if gravity has already started to affect the PDF of the cloud, yielding two extended power-law tails, the variance of $\rho$ can become very large (see \citealt{jaupart2020} and Sect.~(6.3.3) of \citetalias{jaupart2021a}). In that case, the product $\var{\rho/\left<\rho\right>}\times (l_c(\rho)/R)^2  $ can become sizeable and can affect the estimate of $e_G$. 

Indeed in \citet{jaupart2021generalized} the authors show that for a statistically homogeneous density field $\rho$,
\begin{equation}
\esp{\rho} \, \var{\frac{\rho}{\esp{\rho}}} \, l_c(\rho)^3 \simeq \left<\rho\right>\, \var{\frac{\rho}{\left<\rho\right>}} \, l_c(\rho)^3 
\end{equation}
is an invariant of the dynamics of star-forming MCs. Moreover, they show that since this increase in variance due to gravity occurs on a short (local) timescale compared with the typical timescale of variation of $\esp{\rho} \simeq \left<\rho\right>$, the invariant essentially yields
\begin{equation}
 \var{\frac{\rho}{\left<\rho\right>}} \, l_c(\rho)^3 = \mathrm{const},
\end{equation}
at least in the initial phase of gravitational collapse. In that case, as gravity proceeds to concentrate gas in smaller clumpier structures, the product 
\begin{equation}
 \var{\frac{\rho}{\left<\rho\right>}} \, l_c(\rho)^2 (t) \propto    \var{\frac{\rho}{\left<\rho\right>}}^{1/3} (t),
\end{equation}
and as such the contribution of internal structures, can become sizeable since $\var{\rho/\left<\rho\right>}$ becomes very large. Physically speaking, this occurs when gravity has started to break the cloud into small condensed and isolated regions.

\subsubsection{Concluding remarks and the case of Polaris and Orion B}

For typical  initial star-forming  conditions, however, we still  expect $\var{\Sigma/\left<\Sigma\right>}\times (l_c(\rho)/R) \ll 1 $,  as  found in \citetalias{jaupart2021a} for Polaris and Orion B, where $\var{\Sigma/\left<\Sigma\right>}\times (l_c(\rho)/R) \lesssim 10^{-1} $ (see Sect. (6) of \citetalias{jaupart2021a}).

Therefore, we conclude that for typical initial star-forming cloud conditions, the gravitational potential energy can be correctly estimated from the total mass and size of the cloud. The 
(observationally undetermined) internal structure of the cloud yields only a small correction to this average value. Hence, most of the uncertainties comes from the geometrical factors in Eq.~(\ref{eq:eprhomean}). As seen in the next section, however, this is no longer true for simulations in a periodic box of volume $V_{Box} = L^3$.

\subsection{Simulations in a periodic  box} \label{sec:Periodic}

In a periodic box (topology $\mathbb{T}^3$) of volume $V_{Box} = L^3$, the gravitation potential, $\Phi_G$, satisfies the modified Poisson equation:
\begin{equation}
    \Delta \Phi^{L}_{\rm G}(x) = 4 \pi G \left(\rho(x) - \frac{M}{V_{box}} \right),
\end{equation}
where $M$ is the total mass in the box \citep{ricker2008,guillet2011}. This is due to the ill-posed problem of the standard Poisson equation in $\mathbb{T}^3$ 
and is a  well-known feature of statistical mechanics of Coulomb systems. Then, the Green function of the gravitational potential  satisfies
\begin{equation}
   \Delta_x \Phi^{L}_{\rm Green}(x) = 4 \pi G \left( \delta(x) - \frac{1}{V_{box}} \right).
\end{equation}
Rescaling the various fields, $y=x/L$,  one ends up with
\begin{eqnarray}
 \Phi^{L}_{\rm Green}(x) &=&\frac{1}{L}\Phi^{1}_{\rm Green}\left( \frac{x}{L}\right) =  \frac{1}{L}\Phi^{1}_{\rm Green}(y), \\
   \Delta_y \Phi^{1}_{ \rm Green}(y) &=& 4 \pi G (\delta(y) - 1).
\end{eqnarray}
We note that $\Phi^{1}_{\rm Green}$ is periodic (of period $1$) and defined up to a constant which is usually chosen to be such that the average of $\Phi^{1}_{\rm Green}$, or the zero mode of its Fourier transform, is $0$. 
Then 
\begin{equation}
    \left< e_{G} (\left<\rho\right>) \right>_{\mathbb{T}^3} = 0,
\end{equation}
and
\begin{eqnarray}
   \left< e_{G} \right>_{\mathbb{T}^3} &=&  L^2 \int_{[-\frac{1}{2},\frac{1}{2}]^3} \hat{C}_{\rho,L}(L \bm{y}) \, \Phi^{1}_{\rm Green}(y) \mathrm{d}y.
 \end{eqnarray}
Then, in turn,
\begin{equation}
  \left< e_{G} \right>_{\mathbb{T}^3}  =  L^2 \, \mathrm{Var}(\rho) \int_{[-\frac{1}{2},\frac{1}{2}]^3} \Tilde{C}_{\rho,L}(L y) \, \Phi^{1}_{\rm Green}(y) \mathrm{d}y.
\end{equation}
In this case, therefore, the gravitational potential is {\it only a measure of the internal, fluctuating density structure within the box}. 

In the following, we  examine the case of  pure turbulent (initial) conditions (without gravity),  as in \citet{federrath2012,federrath2013}. We thus write
\begin{eqnarray}
   \left< e_{G} \right>_{\mathbb{T}^3} &=& 2 \, G \, \mathrm{Var}(\rho) \, l_{\rm c}^2 \, g_{b,\mathcal{M}} \left( L/l_{\rm c} \right), \label{eq:defgbmach}
\end{eqnarray}
where $g_{b,\mathcal{M}}$ is some bounded dimensionless function that may depend on the Mach number $\mathcal{M}$ and the type of turbulence forcing,  as measured by coefficient $b$. 
Noting that $\mathrm{Var}(\rho) \simeq (b \mathcal{M}) ^2 \left< \rho \right>^2$, we obtain
\begin{eqnarray}
\left< e_{G} \right>_{\mathbb{T}^3} &=& 2 \,G \, (b \mathcal{M}) ^2 \left< \rho \right>^2 \, l_{\rm c}^2 \, g_{b,\mathcal{M}} \left( L/l_{\rm c} \right) \\
&=& 2\, G \, M \frac{\left<\rho\right>}{L} \left(\frac{(b \mathcal{M})l_c}{L}\right)^2 g_{b,\mathcal{M}} \left( L/l_{\rm c} \right).
\end{eqnarray}

 \section{Virial parameter} \label{sec:virial}
 
\begin{figure}
    \centering
    \includegraphics[width=\columnwidth]{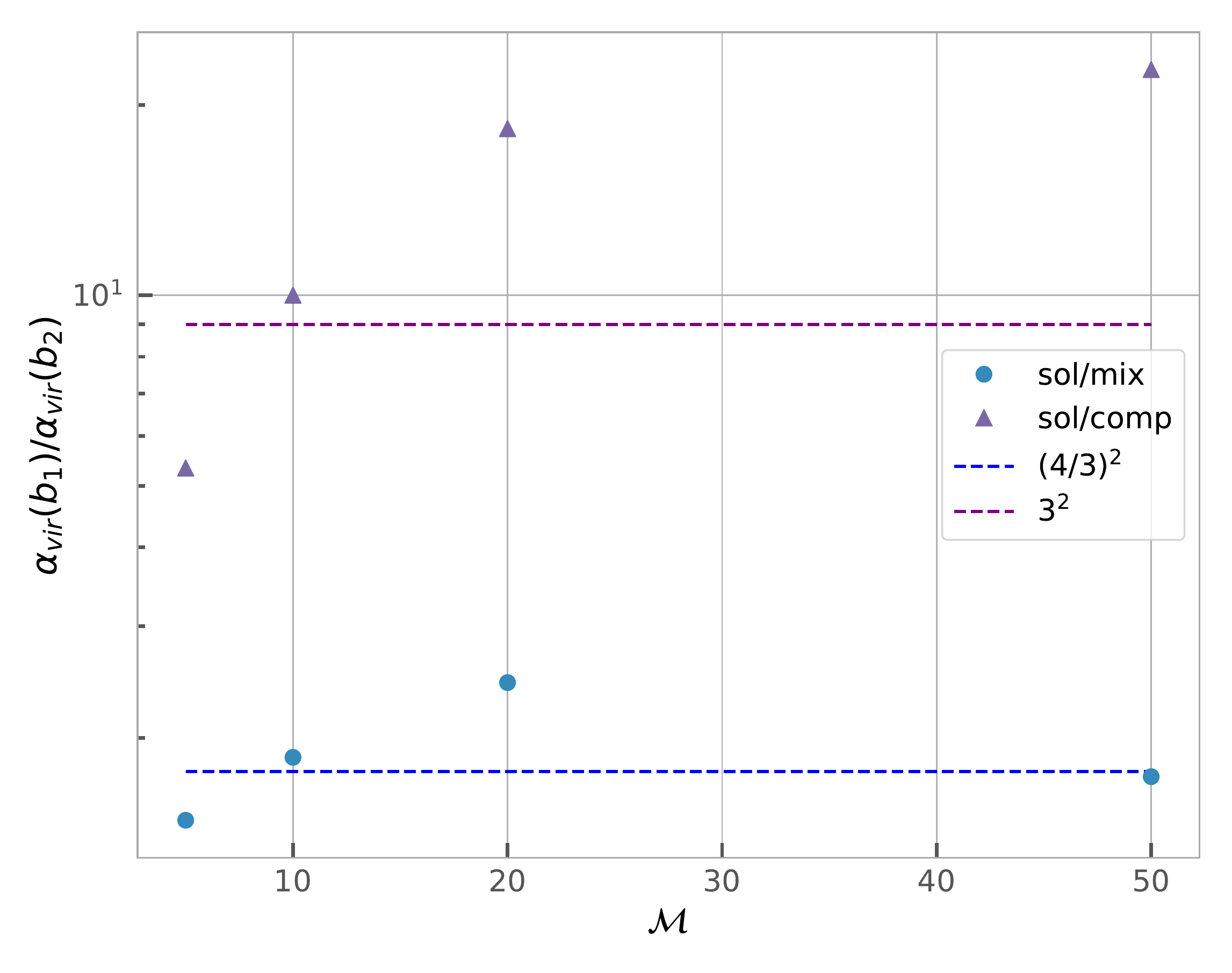}
    \caption{Ratio of virial parameters from the simulations of \citet{federrath2012,federrath2013} for three different types of forcing. Each point corresponds to a Mach number, $\mathcal{M}$, and size $L$. Triangles (sol/comp) show the ratio between virial parameters for solenoidal ($b\simeq0.3$) and compressive forcing ($b\simeq1$) and circles (sol/mix) show the ratio between solenoidal and mixed forcing ($b\simeq0.4$). The horizontal dashed lines give the expected value of th e ratios for a scaling in $b^{-2}$. }
    \label{fig:ratio_virial}
\end{figure}

An important quantity in the study of MCs is the virial parameter, which is defined as the ratio of twice the kinetic energy over the gravitational energy,  
$\alpha_{\rm vir} = 2\left<e_K \right> / |\left<e_G\right>|$ (see e.g. \citealt{mckee1992} for a more complete discussion). 
For clarity purposes, we  again use the simulations of \citet{federrath2012} for comparisons in $\mathbb{T}^3$ and restrict ourselves to isothermal turbulence conditions. 
In these conditions, the following conditional expectation holds \citep{kritsuk2007,federrath2010}:
\begin{equation}
\esp{\mathcal{M}\,|\, \rho } \simeq \esp{\mathcal{M}} , 
\end{equation}
and we get 
\begin{equation}
    \left<e_K\right> \simeq \left<\rho \right> \sigma_V^2 \simeq \esp{\rho} \sigma_V^2,
\end{equation}
where $\sigma_V$ is the 3D velocity dispersion. This yields the virial parameters
\begin{equation}
    \alpha_{\rm vir,\mathbb{R}^3} = \frac{\sigma_V^2 \, L}{2\, G \, M \, (1+\Tilde{\xi_g})\, c_g } =  \frac{\sigma_V^2 \, L}{2\, G \, M} \frac{1}{ (1+\Tilde{\xi_g})\, c_g },
\end{equation}
where $2 \, \Tilde{\xi_g}\, c_g $ is usually taken to be equal to 1 in order to match the virial parameter with that of a homogeneous sphere, and
\begin{equation}
       \alpha_{\rm vir,\mathbb{T}^3} = \frac{\sigma_V^2 \, L}{ 2\, G \, M g_{b,\mathcal{M}} \left( L/l_{\rm c} \right)  } \left( \frac{L}{ (b \mathcal{M}) \, l_{\rm c}}\right)^2 .
\end{equation}
The ratio of the two virial parameters is therefore
\begin{equation}
    \frac{\alpha_{\rm vir,\mathbb{T}^3}}{\alpha_{\rm vir,\mathbb{R}^3}} \simeq \frac{1}{g_{b,\mathcal{M}} \left(L/l_c\right)} \, \left( \frac{L}{ (b \mathcal{M}) \, l_{\rm c}}\right)^2 \label{eq:comparaisonviriel}.
\end{equation}
We can test the validity of this equation. Assuming that the type of turbulence forcing has only a moderate influence on $\Tilde{C}_{\rho, L}$ and thus on $g_{b,\mathcal{M}} \left(L/l_c\right) l_{\rm c}^2$ (see Eq.~(\ref{eq:defgbmach})), 
we have, for a given large-scale Mach number, $\mathcal{M}$, and size $L$:
\begin{equation}
\alpha_{\rm vir,\mathbb{T}^3}/\alpha_{\rm vir,\mathbb{R}^3} \propto b^{-2}. 
\end{equation}
Even though the forcing parameter, $b$, may have some moderate influence on $g_{b,\mathcal{M}} \left(L/l_c\right) l_{\rm c}^2$, 
we still expect the ratio in Eq.~(\ref{eq:comparaisonviriel}) to decrease when  $b$ increases, with a scaling close to $b^{-2}$. 
Indeed, we find a good agreement, within one order of magnitude, between this scaling and the results obtained in \citet{federrath2012,federrath2013}, 
even when we consider variations in the Mach number and virial parameters (Fig. \ref{fig:ratio_virial}).

These calculations demonstrate  the origin of  the large differences between values of the gravitational potential and the virial parameter that are calculated in a  periodic simulation domain ($\mathbb{T}^3$) and those that are derived 
from observations in  space ($\mathbb{R}^3$) from the size and mass of the cloud for pure (initial) turbulent conditions. This mismatch is a numerical artifact due to the numerical resolution of the Poisson equation in a torus geometry. 
It should be emphasized, however, that this does not imply that the numerical results of \citet{federrath2012,federrath2013} are erroneous, but it does call for a reexamination of interpretations 
that involves the estimation of $\alpha_{\rm vir}$ from the gravitational potential returned by the simulations.

\section{Conclusion}

In this article we have applied the statistical formalism developed in \citetalias{jaupart2021a} to the determination of the total gravitational energy and virial parameter of a cloud. We have demonstrated that
the contribution of the (undetermined) internal structure of the clouds has only a small impact on these determinations in clouds  that are not in a too advanced stage of star formation.  This is the case, for example, for the clouds Polaris and Orion B studied in \citetalias{jaupart2021a}. In that case, the cloud gravitational energy and virial parameter can thus be safely estimated from the observed total mass and size, with no knowledge of their internal structures. We, note, however that for clouds within which star formation has already been ongoing for a significant amount of time, that is  when gravity has started to break the clouds into small isolated condensed regions,  contributions from the internal structure can become sizeable and affect the estimate of  the potential energy, $e_G$ (see Sect.~\ref{sec:Isolated}). This is an important result because the virial parameter determines the dynamics of a cloud, equilibrium, expansion or gravitational contraction.

Examining the same problem in a torus geometrical configuration, which is characteristic of numerical simulations in a periodic box, we have shown that, in contrast to real space, only the inner structure of the
density fluctuations in the box contributes to the determination of the  gravitational potential and the virial parameter (see Sect.~\ref{sec:Periodic}). In that case, the (dominant) average contribution is  lacking, a consequence of the  ill-posed problem of solving the Poisson equation in $\mathbb{T}^3$. We have demonstrated that,  for pure (initial) turbulent conditions, the ratio of the viral parameter values in the box over the ones in real geometry is proportional to $\big( (l_{\rm c}/{L})\times (b \mathcal{M}) \big)^{-2}$ (Eq.~(\ref{eq:comparaisonviriel})). Thus, for a given large-scale Mach number and size of the simulation box, this decreases approximately as $\sim b^{-2}$, where $b\in [1/3,1]$ denotes the (solenoidal vs compressive) turbulence forcing parameter.
This explains the puzzling large discrepancy found in \citet{federrath2012,federrath2013} between the gravitational potential and virial parameter values inferred from the global characteristics of the simulation box and those inferred from the numerical results (see Sect.~\ref{sec:virial}). We note that this does not affect the validity of the simulations of \citet{federrath2012,federrath2013}. However, it does call for a reexamination of the interpretations that involve the estimation of $\alpha_{\rm vir}$ from the gravitation potential returned by the simulations.

Finally, we remark that these findings only hold for clouds that are {\it statistically} homogeneous (see Paper I)  and  large enough compared to the correlation length of $\rho$. They notably do not hold for (small-scale) collapsed subregions.

These calculations highlight again the power of the statistical formalism developed in \citetalias{jaupart2021a} to explore the general statistical properties of star-forming MCs from a limited number of observations or simulations. As explored in the present paper, its power is notably significant when estimating its global gravitational energy and virial parameter, and thus the level of binding, of such MCs. These results will be used in a forthcoming paper aimed at exploring the evolution of the PDF in star-forming clouds.

\begin{acknowledgements}
The authors are grateful to Christoph Federrath for always providing data from his numerical simulations upon request.
   \end{acknowledgements}

\bibliographystyle{aa} 
\bibliography{biblio.bib}

\begin{appendix}
\section{Computation of the total potential energy on a control volume $\Omega$.} \label{app:ergodiccalculationgeneralvolume}
We derive the gravitational binding energy of a cloud covering a domain $\Omega$ in Sect.~(\ref{sec:gravpotanygeom}), and divided it into three contributions to isolate the effects of the internal structure (deviation from the average):
\begin{eqnarray}
\left< e_{G} (\left<\rho\right>) \right>  &=& \! \! \int_{\Omega^2} \frac{\left<\rho\right>^2}{|\Omega|} \Phi_{\rm Green}(\bm{x}-\bm{x'}) \, \mathrm{d} \bm{x'} \, \mathrm{d} \bm{x}, \\
  I_C(\rho)\! \! &=& \! \! \int_{\Omega^2} \frac{\delta \rho(\bm{x}) \delta \rho(\bm{x'})}{|\Omega|} \Phi_{\rm Green}(\bm{x}-\bm{x'}) \, \mathrm{d} \bm{x'} \, \mathrm{d} \bm{x},  \\
  2 \left<\rho\right> I_\delta(\rho)  \! \! &=& \! \!  2 \left<\rho\right> \int_{\Omega^2} \frac{\delta \rho(\bm{x})}{|\Omega|} \Phi_{\rm Green}(\bm{x}-\bm{x'}) \, \mathrm{d} \bm{x'} \, \mathrm{d} \bm{x} .
\end{eqnarray}
Then, using the change in variables $(\bm{u},\bm{v})=\varphi(\bm{x},\bm{x'})=(\bm{x}-\bm{x'},\bm{x}+\bm{x'})$, we obtain
\begin{eqnarray}
\left< e_{G} (\left<\rho\right>) \right>  &=& \left<\rho\right> \int_{\varphi_1(\Omega)}  \! \! \mathrm{d} \bm{u} \Phi_{\rm Green}(\bm{u}) \int_{\varphi_2^{\bm{u}}(\Omega)}  \left<\rho\right> \frac{  \mathrm{d} \bm{v}}{8|\Omega|},  \\
I_C(\rho)  \! \! &=& \! \!  \int_{\varphi_1(\Omega)}  \! \! \mathrm{d} \bm{u} \Phi_{\rm Green}(\bm{u}) \int_{\varphi_2^{\bm{u}}(\Omega)}  \frac{ \mathrm{d} \bm{v}}{8|\Omega|}  \delta\rho(\frac{\bm{u} + \bm{v}}{2}) \delta\rho(\frac{\bm{u} - \bm{v}}{2}) \nonumber\\
    &=& \int_{\varphi_1(\Omega)}  \mathrm{d} \bm{u} \Phi_{\rm Green}(\bm{u}) \,  \hat{C}_{\rho,L}(\bm{u}), \\
I_\delta(\rho) \! \!  &=&  \! \! \int_{\varphi_1(\Omega)} \! \!  \mathrm{d} \bm{u} \Phi_{\rm Green}(\bm{u}) \int_{\varphi_2^{\bm{u}}(\Omega)}  \frac{ \mathrm{d} \bm{v}}{8|\Omega|}  \delta\rho(\frac{\bm{u} + \bm{v}}{2}),
\end{eqnarray}
 due to the fact that $\Omega$ possess a center of symmetry and where 
\begin{eqnarray}
\varphi_1(\Omega) &=& 2 \Omega, \\
\varphi_2^{\bm{u}}(\Omega) &=& 2 \left( (\Omega - \bm{u}) \cap \Omega \right) + \bm{u},
\end{eqnarray}
and $\hat{C}_{\rho,L}$ is the biased ergodic estimator of the ACF  of $\rho$  (see Sect.~(2.1.2)  and Appendix (A) of \citetalias{jaupart2021a}). For example, if $\Omega =[-\frac{L}{2},\frac{L}{2}]^3$, then $\varphi_2^{\bm{u}}(\Omega) = \left[-L+ |u_i|, L-|u_i|] \right]$.

Furthermore,  denoting 
\begin{eqnarray}
I(\bm{u}) = \int_{\varphi_2^{\bm{u}}(\Omega)}  \frac{ \mathrm{d} \bm{v}}{8|\Omega|}  \delta\rho(\frac{\bm{u} + \bm{v}}{2}),
\end{eqnarray}
we see that $I(\bm{u})$ is, modulo the factor $1/|\Omega|$, the average of the density deviations, $\delta \rho$, in the sub-volume $\left( (\Omega - \bm{u}) \cap \Omega \right) + \bm{u}/2$. This is easier to see in the pedagogical case where $\Omega =[-\frac{L}{2},\frac{L}{2}]^3$, as
\begin{eqnarray}
I(\bm{u}) \! &=&\! \! \iiint_{-L+|u_i|}^{L-|u_i|}\frac{\mathrm{d}\bm{v}}{8|\Omega|} \delta \rho \left(\frac{\bm{u}+\bm{v}}{2} \right) \nonumber \\
 &=& \frac{1}{|\Omega|} \iiint_{\frac{-L+|u_i|+u_i}{2}}^{\frac{L-|u_i|+u_i}{2}} \delta \rho \left(\bm{x} \right) \, \mathrm{d}\bm{x}. \label{eq:app:neglectedintegral}
\end{eqnarray}
Then, if the volume of $\left( (\Omega - \bm{u}) \cap \Omega \right)$ is sufficiently large,  for example  $|\left( (\Omega - \bm{u}) \cap \Omega \right)| \gg l_c(\rho)^3$, $I(\bm{u}) \simeq |\left( (\Omega - \bm{u}) \cap \Omega \right)| \left<  \delta \rho \right> =0 $ and
\begin{eqnarray}
I(\bm{u}) \ll  \int_{\varphi_2^{\bm{u}}(\Omega)}  \left<\rho\right> \frac{  \mathrm{d} \bm{v}}{8|\Omega|} =  \left<\rho\right>  \frac{ |\left( (\Omega - \bm{u}) \cap \Omega \right)|}{|\Omega|}.
\end{eqnarray}
The integral $I(\bm{u})$ thus only gives non-negligible contributions for $\bm{u}$ in a small volume  of order $l_c(\rho)^3$ near  the border $\partial (2 \Omega)$, such as  $|\left( (\Omega - \bm{u}) \cap \Omega \right)| \lesssim l_c(\rho)^3$. 

Therefore, providing that $L = |\Omega|^{1/3} \gg l_c(\rho)$, we can neglect $ 2 \left<\rho\right> I_\delta(\rho)$ with respect to $\left< e_{G} (\left<\rho\right>) \right>$. This leaves 
\begin{eqnarray}
\left< e_{G} \right> \simeq \left< e_{G} (\left<\rho\right>) \right> + \int_{\varphi_1(\Omega)}  \mathrm{d} \bm{u} \Phi_{\rm Green}(\bm{u}) \,  \hat{C}_{\rho,L}(\bm{u}). 
\end{eqnarray}

\end{appendix}

\end{document}